\documentclass[aps,prl,twocolumn,superscriptaddress,preprintnumbers]{revtex4}
\usepackage{tabularx}
\usepackage{bm}
\usepackage{subfigure}
\usepackage{euscript}
\usepackage{epsfig,psfrag,subfigure}
\usepackage{graphicx,psfrag,subfigure}
\usepackage{color}
\usepackage{amsmath,amsfonts}
\usepackage{exscale}
\usepackage{amsbsy}
\usepackage{subfigure}

\def\be{\begin{equation}}
\def\ee{\end{equation}}
\def\ba{\begin{eqnarray}}
\def\ea{\end{eqnarray}}

\begin{document}

\preprint{INT-PUB-09-024}
\title{Hydrodynamics of chiral liquids and suspensions}
\author{A.~V.~Andreev}
\affiliation{Physics Department, University of Washington, Seattle, WA 98195}
\author{D.~T.~Son}
\affiliation{Institute for Nuclear Theory, University of Washington,
Seattle, WA 98195}
\author{B.~Spivak}
\affiliation{Physics Department, University of Washington, Seattle, WA 98195}
\begin{abstract}
We obtain hydrodynamic equations describing a fluid consisting of
chiral molecules or a suspension of chiral particles in a Newtonian
fluid. The stresses arising in a flowing chiral liquid have a
component forbidden by symmetry in a Newtonian liquid.  For example,
a chiral liquid in a Poiseuille flow between parallel plates exerts
forces on the plates, which are perpendicular to the flow. A generic
flow results in spatial separation of particles of different
chirality. Thus even a racemic suspension will exhibit chiral
properties in a generic flow. A suspension of particles of  random
shape in a Newtonian liquid is described by equations which are
similar to those describing a racemic mixture of chiral particles in
a liquid.
\end{abstract}

\date{May 17, 2009}
\maketitle

Equations of hydrodynamics express conservation of mass, momentum
and energy, and can be written as
\begin{subequations}
\label{HydrEq}
\begin{align}
\partial_t \rho+ \partial_i J_i
&=0, \\
\partial_t P_{i}+ \partial_j \Pi_{ij}&=0,\\
\partial_t E+\partial_i J_E&=0.
\end{align}
\end{subequations}
Here $\partial_t$ and $\partial_i$ denote time and spatial derivatives,
$\rho$, $\mathbf{P}$, and
$E$ are correspondingly the densities of mass, momentum and energy,
and $\mathbf{J}$, $\mathbf{J}_E$ and $\hat \Pi$ are the flux
densities of mass, energy and momentum (we indicate vector
quantities by bold face symbols and second rank tensors by hats).
The flux densities can be expressed in terms of the hydrodynamic
variables: the pressure $p\,(\mathbf{r}, t)$, temperature
$T(\mathbf{r}, t)$ and the hydrodynamic velocity
$\mathbf{v}(\mathbf{r}, t)$, which we define via the equation
\begin{equation}
\rho \mathbf{v}=\mathbf{J} \equiv \mathbf{P}.   \label{hydreq1}
\end{equation}

To lowest order in spatial derivatives we have~\cite{LandauHydr}
\begin{equation}
 \Pi_{ij}= \rho  v_{i}v_{j}+ p\, \delta_{ij}-\eta V_{ij} -
 \zeta \delta_{ij}\, \mathrm{div}\,\mathbf{v}, \label{fluxmomentum}
\end{equation}
where $V_{ij}=\partial_j v_{i}+
 \partial_i
 v_{j}-\frac{2}{3}\delta_{ij}\, \mathrm{div}\,\mathbf{v}$ is the
rate  of shear strain, and  $\eta$ and $\zeta$ are the first and the
second viscosities. This leads to the Navier-Stokes equations, which
should be supplemented by the equation of state of the fluid and the
expression for the energy current in terms of the hydrodynamic variables.

For a dilute suspension of particles in a Newtonian liquid, the
basic hydrodynamic equations need to be
supplemented~\cite{LandauHydr}  by the conservation law for the
current of suspended particles,
\begin{equation}
 \partial_t n + \mathbf{v}\cdot \!\boldsymbol{\nabla} n+  \mathrm{div}
\mathbf{j}=0.
 \label{diffusionequation}
\end{equation}
Here   $n({\bf r}, t)$  is the density of suspended particles, and
$\mathbf{j}(\mathbf{r}, t)$ their flux density  (relative to the
fluid). To linear order in the gradients of concentration,
temperature and pressure the latter can be written as
\begin{equation}\label{eq:diff_current}
    \mathbf{j}=-D \boldsymbol{\nabla} n- n \lambda_T \boldsymbol{\nabla} T -n
\lambda_p \boldsymbol{\nabla} p.
\end{equation}
Equation (\ref{hydreq1}) remains unchanged and can be considered as
a definition of the hydrodynamic velocity ${\bf v}$, which is,
generally speaking, different from the local velocity ${\bf u}( {\bf
r},t)$ near an individual particle of the suspension.

There are corrections to the flux densities of various quantities,
which are higher orders in spatial derivatives of the hydrodynamic
variables (for a review see, for example,
Refs.~\onlinecite{GalkinFrindler,LandauKin}). Moreover, there are
nonlocal corrections to the Navier-Stokes equations, which can not
be expressed in terms of higher order spatial derivatives of
hydrodynamic variables\cite{AW,AW1,EHvL,Andreev}.

Several studies focused on the effects of chirality on the
motion of suspended particles in hydrodynamic
flows~\cite{DeGennes,Brand,Doi1,Doi_ribbons,Doi2,Kostur,Chen2007,Witten,
Marcos2009}. It was shown that non-chiral magnetic
colloidal particles can self-assemble into chiral colloidal
clusters~\cite{Chaikin}.

In this article, we develop a hydrodynamic description for the case
of a suspension containing both right-handed and left-handed chiral
particles in a centrosymmetric liquid.  We show that in this case
the corrections to the Navier-Stokes equations contain new terms,
which are associated with the chirality of the particles. The
significance of these corrections is that they describe new effects,
which are absent in the case of centrosymmetric liquid. Since
certain types of hydrodynamic flow lead to separation of particles
with different chirality, these corrections are important even in
initially racemic suspensions of chiral particles. For simplicity we
consider the case of incompressible fluids.

In a given flow an individual particle of the suspension undergoes a
complicated motion which depends on the initial position and
orientation of the particle.  The hydrodynamic equations can be
written for quantities which are averaged over the characteristic
spatial and temporal scales of such motion.

In the presence of chirality the following contribution to momentum flux
density is allowed by symmetry:
\begin{equation}\label{correctionmomentflux}
\Pi^{ch}_{ij}=n^{ch}\alpha \eta [\partial_i\mathbf{\omega}_{j} +
\partial_j\mathbf{\omega}_{i}] + \eta \alpha_1  [\omega_i \partial_j n^{ch}  +
\omega_j \partial_i n^{ch} ],
\end{equation}
where $\omega_i(\mathbf{r})=\frac{1}{2}\, \epsilon_{ijk}\partial_j v_k
(\mathbf{r})$ is the flow vorticity, and $n^{ch}=(n_{+}-n_{-}) $ is the chiral
density, with
$n_{+}$ and $n_{-}$ being the volume densities of right- and left-handed
particles
respectively.
Equations (\ref{HydrEq}-\ref{correctionmomentflux}) should be
supplemented by the expression for the chiral current, defined as
the difference between the currents of right- and left-handed
particles. Separating it into the convective part,
$\mathbf{v}n^{ch}$, and the current relative to the fluid,
$\mathbf{j}^{\,ch}$, we write the continuity equation as
\begin{equation}
\partial_t n_{ch} + \mathrm{div}(\mathbf{v} n^{ch}) +
\mathrm{div} \mathbf{j}^{\,ch}=0. \label{chiraldiffusion}
\end{equation}

Besides the conventional contribution given by Eq.~(\ref{eq:diff_current}) with
$n$ replaced by $n^{ch}$, the chiral current contains a
contribution,
$\tilde{\mathbf{j}}^{\, ch}$, which depends on the flow vorticity:
\begin{eqnarray}
\tilde{j}^{ch}_{i}&=&n [\beta \nabla^2 \omega_{i} + \beta_{1}
\omega_{j}V_{ij}], \label{chiralflux}
\end{eqnarray}
where $n=n_{+}+n_{-}$.  The contributions to $\tilde{j}^{\,ch}_i$
containing only $n\omega_i$ are not allowed as there should be no
chiral current in rigidly rotating fluid.

In Eqs.~(\ref{correctionmomentflux}) and (\ref{chiralflux}) we keep
only the leading terms in the powers of $\partial_i v_j$, or in the
order of spatial derivatives of $\mathbf{v}$. Although these terms
are subleading in comparison to those in the conventional
hydrodynamic approximation,  they describe new effects which are
absent in the latter.

Note that according to the Navier-Stokes equations
\begin{equation}\label{eq:relation}
\nabla^2 \mathbf{curl} \, ({\bf v})=\frac{\rho}{\eta}\{ \partial_t
\mathbf{curl} \, ({\bf v})+\mathbf{curl}\, [({\bf v}\nabla){\bf
v}]\}.
\end{equation}
Thus the first term in Eq.~(\ref{chiralflux}) arises either due to
non-stationary or non-linear in ${\bf v}$ nature of the flow. In
particular, in  stationary flows and to zeroth order in the Reynolds
number $\nabla^2 \mathbf{curl}(\mathbf{v})=0$ and this term
vanishes.

In spatially inhomogeneous flows the suspended particles rotate,
generally speaking, relative to the surrounding fluid. This gives
rise to separation of particles of different chirality due to the
propeller effect, and to the chiral contribution to the momentum
flux, Eq.~(\ref{correctionmomentflux}).

The rotation of the particles relative to the fluid arises due to
two effects:

i) In the presence of the spatial dependence of vorticity, $\omega_i
(\mathbf{r})$, the angular velocity of a particle is different from
$\omega_i(\mathbf{r})$. This results in
Eq.~(\ref{correctionmomentflux}) and the first term in
Eq.~(\ref{chiralflux}).

ii) A non-uniform hydrodynamic flow induces orientational order in
suspended particles similar to nematic order in liquid crystals. In
the presence of flow vorticity orientation
of particles induces their rotation with respect to the surrounding
fluid. This contributes both to the chiral stress and
the chiral flux. The latter contribution is
described by the second term in Eq.~(\ref{chiralflux}). The
contribution to the chiral part of the stress tensor associated with
orientational order was discussed in Ref.~\onlinecite{Brand}.

In most cases of practical importance the Reynolds number
corresponding to the particle size $R$ is small. In this regime the
coefficients $\alpha$, $\alpha_1$, $\beta$, and $\beta_{1}$ in
Eqs.~(\ref{correctionmomentflux}) and (\ref{chiralflux}) can be
obtained by studying the particle motion in the surrounding fluid in
the creeping flow approximation~\cite{Kim,Brenner}. In this
approximation the motion of particle immersed in the liquid is of
purely geometrical nature (see for example
Ref.~\onlinecite{Wilczek}). Dimensional analysis gives an estimate
\begin{equation}\label{eq:alpha_estimate}
\alpha\sim \alpha_1 \sim \chi R^4, \quad \beta\sim   \chi R^{3},
\end{equation}
where $R$ is the characteristic size of the particles, and the
dimensionless parameter $\chi$ characterizes the degree of chirality
in shape of the particles.

The relative magnitudes of the different terms in
Eqs.~(\ref{correctionmomentflux}) and (\ref{chiralflux})  depend on
the particle geometry. For example,  particles with the symmetry of
the isotropic helicoid~\cite{Brenner} can not be oriented in a shear
flow. Therefore the second term in Eq.~(\ref{correctionmomentflux})
vanishes is this case.

The degree of orientation of the particles can be obtained by
balancing the characteristic directional relaxation rate due to the
Brownian rotary motion, $\sim T/\eta R^3$ with $T$ being the
temperature, with the rate of orientation due to the shear flow,
$\sim V_{ij}$.  Thus at small shear rates the degree of particle
orientation is $\sim V_{ij} \eta R^3/T $. This leads to the estimate
\begin{equation}\label{eq:orientation_estimate}
\beta_1 \sim \chi \eta R^4 /T.
\end{equation}
The second term in Eq.~(\ref{chiralflux}) is the leading term in the
expansion in the rotational P\'eclet number $\mathrm{Pe} \sim
V_{ij}\eta R^3/T$. At larger P\'eclet numbers terms of higher power
in $V_{ij}$ should be taken into account. For $\mathrm{Pe} \gg 1$
the particle orientation becomes strong, and the corresponding
contribution to the chiral current can be estimated as
\begin{equation}\label{eq:Pe_infty}
    \tilde j^{ch}\sim \chi R \, \omega.
\end{equation}

Equations (\ref{correctionmomentflux}) and (\ref{chiralflux}) are
written for the case when there is no external force acting on the
particles, e.g. for a suspension of uncharged neutrally buoyant
particles. In the presence of an external force $\mathbf{F}$, there
will be additional contributions to the chiral flux.  The linear in
$\mathbf{F}$  contributions can be constructed by contracting the
antisymmetric tensor $\epsilon_{ijk}$ with the velocity $v_i$, force
$F_i$ and two derivatives $\partial_i$.  For example, the following
terms exist when ${\bf F}$ is constant: $({\bf
F}\cdot\bm{\nabla})\bm{\omega}$, ${\bf F}\times\nabla^2{\bf v}$,
$\bm{\nabla}({\bf F}\cdot\bm{\omega})$. These terms arise when the
orientation of the suspended particles can be characterized by a
polar vector. In this case the degree of particle orientation can be
estimated as $ \sim R F/T$. Thus the coefficients with which these
terms enter the chiral current ${\bf j}^{\,ch}$ are of order as
$\chi R^{3}/T$. In the case when particles do not have a polar axis
the degree of their orientation, and the corresponding contribution
to the chiral flux are quadratic in ${\bf F}$ for small force.

\begin{figure}[ptb]
\includegraphics[width=8.0cm]{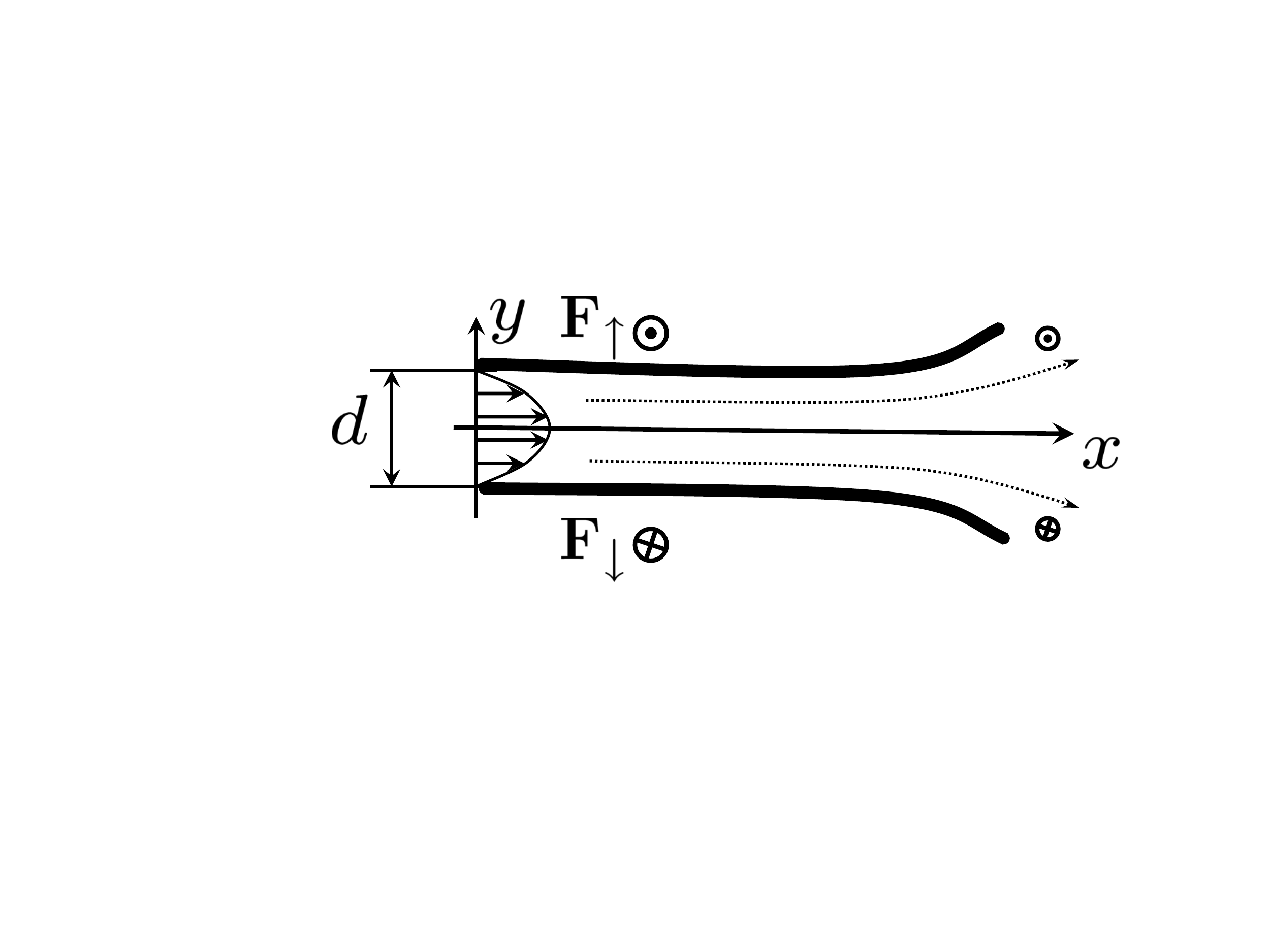}
\caption{A chiral liquid in a Poiseuille flow between parallel
plates exerts a pair of opposite forces on the plates
$\mathbf{F}_\uparrow =-\mathbf{F}_\downarrow$, which are directed
into ($\otimes$) and out of ($\odot$) the page. A flow of a chiral
liquid in a converging or diverging channel develops a helicoidal
component of velocity directed into and out of the page, as shown at
right. For a Newtonian fluid the flow lines (dotted lines) lie in
the plane of the figure.} \label{fig:flows}
\end{figure}

The chiral contribution to the stress tensor
Eq.~(\ref{correctionmomentflux}) leads to several new effects.
Consider a Poiseuille flow of a chiral liquid between parallel
planes separated by a distance $d$: $v_x= -\partial_x p
(d^2-4y^2)/8\eta$, $v_z=v_y=0$ (see Fig.~\ref{fig:flows}), with
$\partial_x p$ being the pressure gradient along the flow.  If the
chiral density is uniform the chiral part of the stress tensor has
only two non-vanishing elements, $\Pi_{yz}=\Pi_{zy}=-\alpha
n^{ch}\partial_x p/2$. It describes a pair of opposing forces per
unit area exerted by the liquid on the top and bottom planes. These
forces are perpendicular to the flow, as shown in
Fig.~\ref{fig:flows}. Assuming $n^{ch} R^3\sim 1$ and using
Eq.~(\ref{eq:alpha_estimate}) the magnitude of the chiral force per
unit area of the plane can be estimated as $\sim \chi R \partial_x
p$.

If $n^{ch}$ is constant in space, then the volume force density
generated by the chiral part of the stress tensor is
$f^{ch}_i=-\partial_j \Pi^{ch}_{ij}=-n^{ch} \alpha \eta \nabla^2
\omega_i$. Then it is clear from Eq.~(\ref{eq:relation}) that
$f^{ch}_i$ is generated only in nonstationary or nonlinear flows. In
the special case of stationary Poiseuille flow the chiral part of
the stress tensor does not generate a force density inside the fluid
even at large Reynolds numbers. Thus the flow pattern is not
affected by the fluid chirality. However, in a generic flow with
converging or diverging flow lines the fluid chirality \emph{does}
affect the flow pattern. This is especially evident in flows, which
have a mirror symmetry in the absence of chiral corrections. In
these cases the chiral contribution to the stress tensor results in
mirror asymmetric corrections to the flow velocity. For example a
chiral liquid flowing between two surfaces with a varying distance
between them, see Fig.~\ref{fig:flows}, will develop a helicoidal
component of velocity with non-vanishing vorticity along the flow
direction. This can be checked explicitly for the exactly solvable
flow in a converging channel ($\S$ 23 of
Ref.~\onlinecite{LandauHydr}).

Another consequence of Eq.~(\ref{chiralflux}) is a possibility of
separation of particles of different chirality in hydrodynamic
flows. It has been observed in numerical
simulations~\cite{Doi1,Doi_ribbons,Doi2,Kostur} and recent
experiments~\cite{Marcos2009}. We note that according to
Eq.~(\ref{eq:relation}) in a stationary flow and in the linear
approximation in the shear rate $\partial_{i}v_{j}$, we have
$\nabla^{2} \omega_i=0$, and the first term in
Eq.~(\ref{chiralflux}) vanishes. Thus separation chiral isomers in
the absence of external forces  acting on the particles is possible
either in non-linear or in non-stationary flows.

In the practically important case of a stationary Couette flow, the
first term in Eq.~(\ref{chiralflux}) vanishes for arbitrary Reynolds
numbers, and the chiral current arises only due to orientation of
the particles. The latter increases with the rotational P\'eclet
number and saturates at $\mathrm{Pe}\gg 1$. In this regime the
chiral current becomes linear in the flow vorticity $\omega$,
Eq.~(\ref{eq:Pe_infty}). The linear dependence of ${\bf j}_{ch}$
on $\omega$ and saturation of the proportionality coefficient at
$\mathrm{Pe} \to \infty$ has been observed numerically in
Refs.~\onlinecite{Doi1,Doi2}.

Separation of particles by chirality can also be achieved by
subjecting the particles to an external circularly polarized
electric or magnetic field. The orientation of the particles along
the field (e.g. due to the presence of a permanent  electric or
magnetic dipole moment or anisotropy of the polarization matrix)
will cause their rotation relative to the surrounding fluid. This
will produce a chiral flux along the circular polarization axis due
to the propeller effect~\cite{Zeldovich,Pomeau} (similarly, a
stationary electric or magnetic field will cause separation of
particles by chirality in a rotating fluid). We also note that there
are other mechanisms of chiral current generation which do not
involve transfer of angular or linear momentum from the ac-field to
the particles~\cite{Spivak2009}. Chiral separation by circularly
polarized magnetic field has been recently observed in the
experiments of Refs.~\onlinecite{Zhang2009,Ghosh2009}. The full
quantitative analysis of this effect is beyond the scope of this
work. Here we restrict the treatment to the experimentally relevant
regime of strong and slowly varying fields,  where thermal
fluctuations can be neglected and the particles are fully polarized
along the instantaneous electric field. In this case  the problem is
of purely geometric nature. The chiral current becomes independent
of the viscosity of the fluid and can be expressed in terms of the
Berry adiabatic connection~\cite{Wilczek}. Below we express this
adiabatic connection in terms of the resistance matrix of the
particle~\cite{Brenner}. The latter relates the external force
$\mathbf{F}$ and torque $\boldsymbol{\tau}$ exerted on the particle
to the linear velocity $\delta \mathbf{v}$ and angular velocity
$\delta \boldsymbol{\omega}$ relative to the fluid,
\begin{equation}\label{eq:resstance_matrix}
    \left(
      \begin{array}{c}
        \mathbf{F} \\
        \boldsymbol{\tau} \\
      \end{array}
    \right)= -
    \left(
      \begin{array}{cc}
        \hat{K} & \hat{C} \\
        \hat{C} & \hat{\Omega} \\
      \end{array}
    \right)\left(
             \begin{array}{c}
                \delta \mathbf{v}\\
               \delta \boldsymbol{\omega} \\
             \end{array}
           \right).
\end{equation}
Here we chose the origin of the reference frame at the reaction
center, so that the coupling tensor $\hat{C}$ is
symmetric~\cite{Brenner}. Since a uniform electric field exerts no
force on the particle we immediately obtain the relations,
\begin{subequations}
\begin{eqnarray}
 \delta\mathbf{v}&=&-\hat{K}^{-1}\hat{C}\delta \boldsymbol{\omega},
  \label{eq:velocity_omega}\\
 \boldsymbol{\tau}&=&-\tilde{\Omega} \delta \boldsymbol{\omega},\label{taudom}
\end{eqnarray}
\end{subequations}
where we introduced the notation $\tilde{\Omega}= \hat{\Omega}-\hat
C \hat K^{-1} \hat C$.

The orientation of the particle relative to the lab frame is
described by the three Euler angles, $\phi$, $\theta$ and
$\psi$~\cite{LandauMech}. We choose the axes of the body reference
frame, $x_1,x_2,x_3$, so that $x_3$ points along the dipole moment
of the particle. For fully polarized particles the latter points
along the instantaneous direction of the electric field. Thus
$\theta$ and $\phi$ coincide with the polar angles of the electric
field vector. The value of $\psi$ remains undetermined because the
particle can be rotated by an arbitrary angle about $x_3$ without
changing its polarization energy. When the orientation of the
electric field changes with time the particle orientation angle
about the instantaneous field direction, $\psi(t)$, also changes.
Its time evolution can be determined from the condition that
projection of the torque onto the $x_3$ axis must vanish. This is
clear because the torque $\boldsymbol{\tau}=\mathbf{d}\times
\mathbf{E}$ is perpendicular to the dipole moment $\mathbf{d}$,
which points along $x_3$. Writing Eq.~(\ref{taudom}) in the body
frame, $\tilde\Omega_{3i}\,\omega_i=0$, and expressing $\omega_i$ in
terms the Euler angles, $\omega_1 = \dot\phi \sin \theta \sin \psi
+\dot\theta \cos\psi$, etc. we obtain $d \psi = A_\phi d\phi
+A_\theta d\theta$ where $A_\phi$  and $A_\theta$ play the role of
the adiabatic connection components and are given by
\begin{subequations}
\label{eq:vector_potential}
\begin{align}
  A_\phi &= -\cos \theta -\frac{\sin\theta}{\tilde{\Omega}_{33}}\left(
  \tilde{\Omega}_{31}\sin\psi + \tilde{\Omega}_{32}\cos\psi \right),  \\
  A_\theta &=-\frac{1}{\tilde{\Omega}_{31}}\left(
  \tilde{\Omega}_{31}\cos\psi + \tilde{\Omega}_{32}\sin\psi \right).
\end{align}
\end{subequations}
This defines $\dot\psi$ in terms of $\dot\theta$ and $\dot\phi$. The
displacement of the particle can be obtained from
Eq.~(\ref{eq:velocity_omega}). Instead of presenting the general
formulae we focus on the practically relevant case of an ac-field of
frequency $\omega_0$ circularly polarized in the $xy$ plane:
$\theta=\pi/2$, $\phi=\omega_0 t$. In this case it is easy to see
that the average velocity along the $x$ and $y$ axes vanishes. For
the average $z$-component of the velocity an elementary calculation
gives,
\begin{equation}\label{eq:velocity}
    \delta v_z=\chi R \omega_{0} =\frac{\omega_0}{2\tilde\Omega_{33}}\mathrm{Tr}\left[
\hat{K}_{-1} \hat{C} \left(
    \begin{array}{ccc}
       -\tilde\Omega_{33} & 0 & 0 \\
       0 & -\tilde\Omega_{33} & 0 \\
       \tilde\Omega_{31} & \tilde\Omega_{32} & 0 \\
    \end{array}
    \right)
     \right],
\end{equation}
where the resistance tensor is expressed in the body frame, and
$\chi$ can be viewed as the dimensionless measure of the particle
chirality. Since the coupling tensor $\hat{C}$ changes sign under
inversion it is clear that particles of opposite chirality will move
in opposite directions along the $z$ axis. By order of magnitude the
chiral separation velocity is $v^{\,ch}_{z}\sim R\omega_0$, which is
consistent with the recent experimental
findings~\cite{Zhang2009,Ghosh2009}.

In the regime where the polarization energy in the electric field is
smaller than the temperature the chiral current is reduced compared
to the above estimate. The leading contribution at weak fields is
proportional to the intensity of the ac-radiation~\cite{Zeldovich}.

So far we discussed the case where the suspended particles consist
of the opposite enantiomers of a single species. However, the
effects considered above exist even in suspensions of particles of
completely random shape in a non-chiral liquid. In this case the
definition of chirality requires clarification. For example one can
define the chirality of a particle by considering the direction of
its motion in a hydrodynamic shear flow or under the action of an
ac-electromagnetic field. Thus the same individual particle can
exhibit different chirality with respect to different external
perturbations.

The set of Eqs.~(\ref{HydrEq}-\ref{chiralflux}) still holds for a
suspension of random particles. In this case the auxiliary
quantities $n^{ch}$, and $\tilde{\bf j}^{\,ch}$ are defined in terms
of the chiral component of the stress tensor
Eq.~(\ref{correctionmomentflux}) and correspond to quantities
averaged over the random shape of the particles.

Finally, we note that symmetry allows contributions to ${\bf
j}_{ch}$ that are proportional to the external magnetic field ${\bf
B}$, for example $\tilde{\mathbf{j}}^{ch}\propto n^{ch}(\nabla
T)^{2}{\bf B}$. We believe that these effects are of fluctuational
origin similar to those discussed in
Refs.~\onlinecite{AW,AW1,EHvL,Andreev} and do not study them in this
work.

We acknowledge useful discussions with D. Cobden, E. L. Ivchenko and
L. Sorensen. This work was supported by the DOE grants
DE-FG02-07ER46452 (A.V.A.), DE-FG02-00ER41132 (D.T.S.) and  by the
NSF grant DMR-0704151 (B.S.).

\end{document}